\documentclass[cits]{PoS}

\newcommand{\Rs}{\ensuremath{R_{\odot}}}
\newcommand{\Ms}{\ensuremath{M_{\odot}}}
\newcommand{\ie}{{\it i.e.}}
\newcommand{\eg}{{\it e.g.}}
\newcommand{\viz}{{\it viz.}}
\newcommand{\ga}%
        {\,\hbox{\lower0.6ex\hbox{$\sim$}\llap{\raise0.6ex\hbox{$>$}}}\,}
\newcommand{\la}%
        {\,\hbox{\lower0.6ex\hbox{$\sim$}\llap{\raise0.6ex\hbox{$<$}}}\,}

\title{Stellar mass black holes in star clusters: gravitational wave emission and detection rates}

\ShortTitle{Stellar mass black holes in star clusters}

\author{\speaker{Sambaran Banerjee}\\
        Argelander-Institut f\"ur Astronomie, University of Bonn, Germany\\
        E-mail: \email{sambaran@astro.uni-bonn.de}}

\abstract{
We investigate the dynamics of stellar-mass black holes (BH) in star clusters
focusing on the dynamical formation of BH-BH binaries, which are
very important sources of gravitational waves (GW). We examine
the properties of these BH-BH binaries through direct N-body computations of
Plummer clusters, having initially $N(0) \leq 10^5$ low mass stars and a population
of stellar mass BHs, using the state-of-the-art N-body integrator ``NBODY6''.
We find that the stellar mass BHs
segregate rapidly into the cluster core to form a central dense sub-cluster of BHs 
in which BH-BH binaries form via 3-body encounters. While most of the
BH binaries finally escape from the cluster
by recoils due to super-elastic encounters with the single BHs,
we find that for clusters with $N(0) \ga 5\times 10^4$, typically
a few of them dynamically harden to the extent that they can merge via GW emission
within the cluster. Also for each of such clusters,
there are a few escaped BH binaries that merge
within a Hubble time, most of the merger happening within a few Gyr of cluster evolution.
These results imply that the intermediate-aged massive clusters
constitute the most important class of star clusters
that can produce dynamical BH-BH mergers at the present epoch.
The BH-BH merger rates obtained from our computations imply a significant
detection rate ($\approx 30{\rm ~yr}^{-1}$) for the ``Advanced LIGO''
GW detector that will become operative in the near future.
Finally, we briefly discuss our ongoing development on this work incorporating the formation of BHs in
star clusters from stellar evolution. In particular, we highlight on the effect of
stellar metallicity on the BH sub-cluster driven expansion of a star cluster's
core.
}

\FullConference{25th Texas Symposium on Relativistic Astrophysics - TEXAS 2010\\
		December 06-10, 2010\\
		Heidelberg, Germany}

\begin{document}

\section{Introduction}\label{intro}

Star clusters, \eg, globular clusters (henceforth GC), young and intermediate-age massive
clusters harbor a large overdensity of compact stellar remnants compared to that
in the field by virtue of their higher stellar density and the mass segregation of the stellar remnants.
These compact stars, which are neutron stars
(hereafter NS) and black holes (hereafter BH), are produced through the evolution of the
most massive stars in the cluster within the first $\sim$ 50 Myr after cluster formation.
Since the BHs are significantly heavier than the remaining low-mass stars of the cluster,
they undergo a runaway mass-segregation or mass-stratification instability \cite{spz} provided a significant
fraction of them are retained in the cluster following their progenitor supernovae.
Due to this instability (also known as the Spitzer instability), the BHs concentrate quickly to the cluster core
forming a dense sub-cluster made purely of BHs which is nearly dynamically isolated from rest of
the cluster. 

Earlier studies of the dynamical properties of such BH sub-clusters 
(\eg, references \cite{olr2006,mak2007,bbk2010}) indicate that
the density of the BHs there is large enough that the BH-BH binary formation
through 3-body encounters \cite{hh2003} becomes important. These dynamically
formed BH binaries then shrink in orbit or ``harden'' through repeated super-elastic encounters
with the surrounding BHs (Heggie's law \cite{h75}).
The binding energy of the BH binaries released in this way is
carried by the single BHs and BH binaries involved in the encounters. This
causes the single and the binary BHs to get ejected from the BH core
to the outer regions of the cluster and they deposit their energy to the cluster stars while returning
to the core via dynamical friction \cite{spz,hh2003}, thereby heating the cluster. This heating mechanism
is most efficient in the central region of the cluster since the stellar density is there the highest.
As the BH binaries harden, the encounter-driven recoils in the BH core become increasingly stronger and finally
the recoils are large enough that the encountering single BHs and/or the BH binaries
escape from the cluster. This also results in heating of the cluster core
as the associated mass-loss dilutes the potential well there.
These heating mechanisms result in an expansion of the central part of the cluster and
modify the dynamical evolution of the cluster significantly.

In the present work, we make a detailed study of
the dynamics of the BH-BH binaries formed in a BH sub-cluster.
Specifically, we investigate whether hard enough BH binaries,
that can merge via gravitational radiation in a Hubble time
within the cluster or after being ejected, can be formed
in such a sub-cluster. To that end, we perform a series of direct N-body integrations of concentrated
star clusters (half-mass radius $r_h \leq 1$ pc) consisting of 
$N \sim 10^4 - 10^5$ low-mass stars, in which a certain number of stellar-mass BHs
are added, representing a star cluster with an evolved stellar population.
However, we also consider initial clusters with a full stellar mass spectrum
in which the BHs are formed from stellar evolution.

\section{Computations}\label{sim}

\begin{figure}
\centering
\includegraphics[width=7.4cm,angle=0]{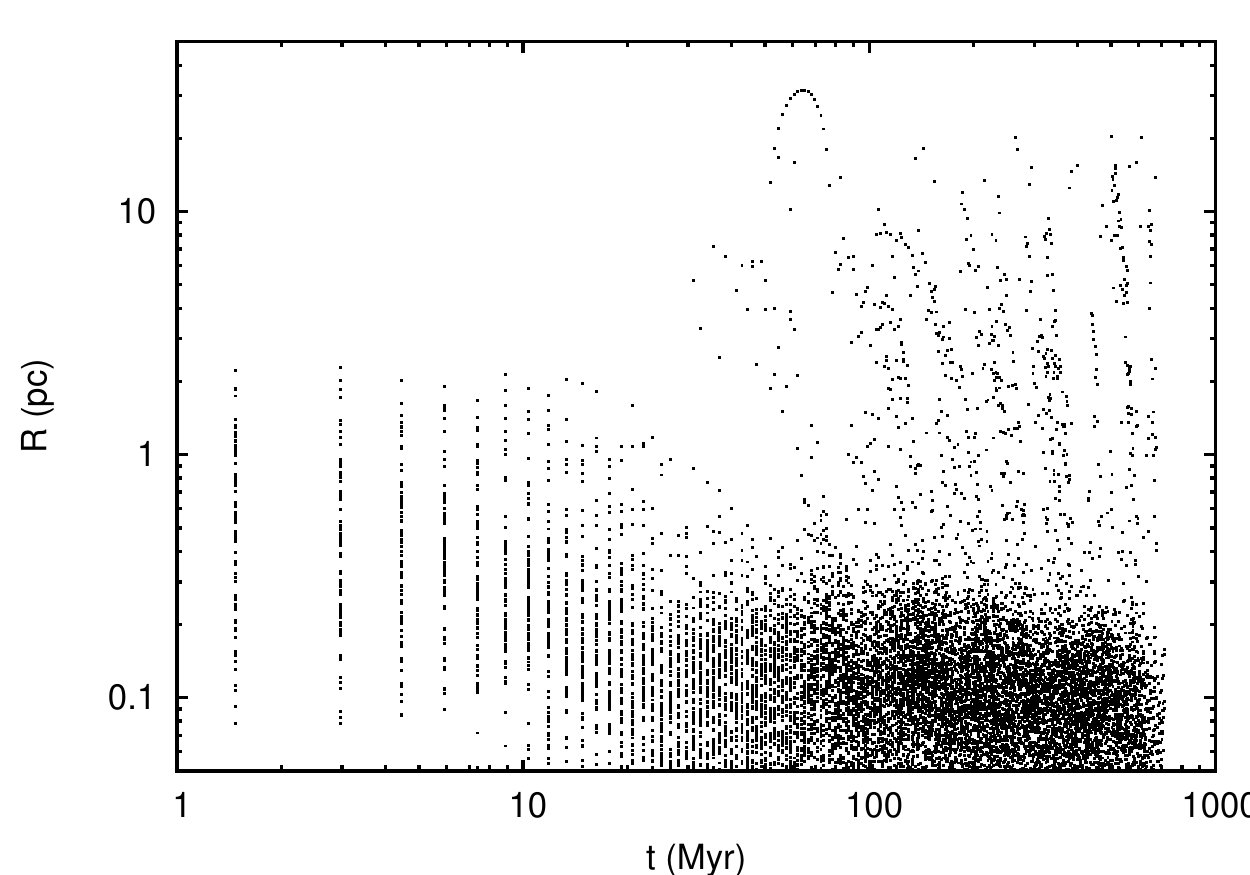}
\includegraphics[width=7.4cm,angle=0]{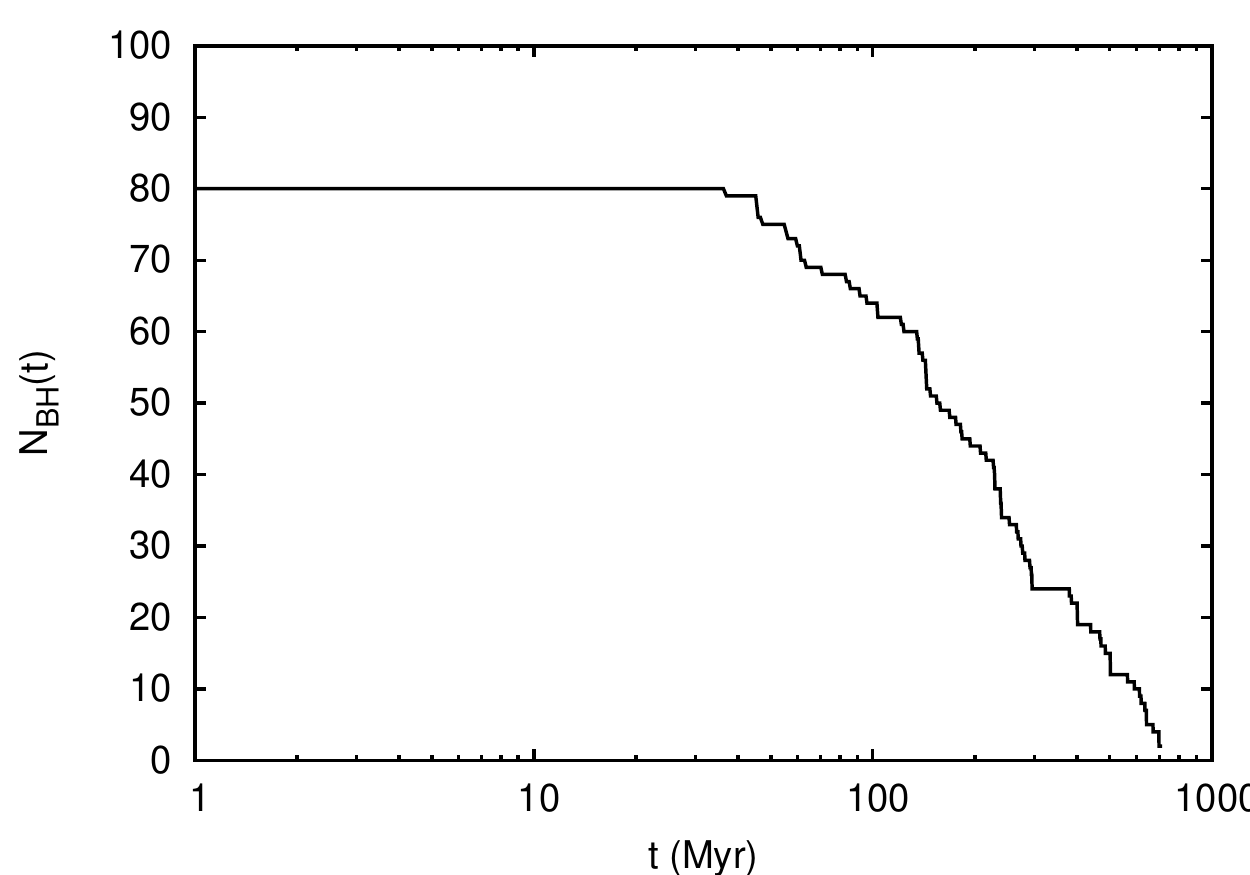}
\caption{Mass segregation of BHs shown by the
radial position $R$ vs. time $t$ (left panel) for the model C50K80.
The BHs segregate in $\sim 50$ Myr during
which $N_{BH}$ (the number of BHs bound to the cluster) remains constant (right panel).
As the BH sub-cluster becomes dense enough that BH-BH binaries begin to form
via 3-body encounters, single BHs and BH binaries start escaping
from the cluster depleting $N_{BH}$.}
\label{fig:bhpos}
\end{figure}

To study the dynamics of BHs in star clusters, we perform direct N-body
integrations of model star clusters using the ``NBODY6'' N-body integration code running
on Graphical Processing Unit (GPU) platforms \cite{arhp}. The initial clusters follow a Plummer model with
half-mass radius $r_h(0) \leq 1$ pc consisting of $N(0) \leq 10^5$
low-mass main-sequence stars in the mass-range $0.5\Ms \leq m \leq 1.0\Ms$.
A specified number of BHs are added to each cluster with the same initial Plummer distribution as
the stars. The initial number of BHs are chosen according to the BH retention fraction,
where we explore both the full retention and the case where half of the BHs have
escaped from the cluster by supernova natal kicks. For simplicity, we consider only equal-mass BHs
in the present work, with the representative value of $M_{BH} = 10\Ms$.
Such an initial condition mimics the epoch at which the massive stars have already been evolved to
produce their remnant BHs.

To evolve the BH-BH binaries due to GW radiation, the well known Peters'
formula is adapted in NBODY6, according to which
the merger time $T_{mrg}$ of an equal-mass BH-BH binary due to gravitational wave (GW) emission is given by,
\begin{equation}
T_{mrg} = 150{\rm Myr}\left(\frac{\Ms}{M_{BH}}\right)^3
\left(\frac{a}{\Rs}\right)^4(1-e^2)^{7/2}, 
\label{eq:tmrg}
\end{equation}
where, $a$ is the semi-major-axis of the binary and $e$ is its eccentricity. 
Studies in numerical relativity indicate that for unequal-mass BH-BH mergers
or even for equal-mass BHs with unequal spins, the merged BH product acquires
a velocity-kick of typically 100 km s$^{-1}$ or more due to the asymmetric momentum outflow
from the system, associated with the asymmetry of the GW emission.
Therefore, in our computations, we provide an arbitrarily large velocity kick of
150 km s$^{-1}$ to the merged BH immediately after a BH-BH merger occurs within the cluster,
to make sure that it escapes.
Also, we evolve the clusters isolated, \ie, in absence of a galactic tidal field,
as the formation of the BH-core through mass segregation and its dynamics
remain largely unaffected by the presence of a tidal field, which
mainly influences the stars near the tidal boundary. Furthermore,
for simplicity, we do not consider primordial binaries
in this initial study. See reference \cite{bbk2010} for the details on the implementation
of initial conditions and GW natal kicks.

\subsection{Mimicking a GC core: reflective boundary}\label{reflct}

We also perform computations with a smaller number of stars and BHs that are confined within
a reflecting spherical boundary \cite{bbk2010}. With such a dynamical system,
one can mimic the core of a massive cluster, where the BHs are concentrated
after mass segregation. The advantage of this approach is that one can compute
the evolution of a massive cluster with many fewer stars, allowing
much faster computation. We integrate
$N = 3000 - 4000$ stars bounded within $0.4$ pc which gives
a stellar density of $\sim 10^4 \Ms {\rm ~pc}^{-3}$ appropriate for the
core-density of a massive cluster. However, stars and BHs faster than
a pre-assigned speed $v_{esc} \approx 24 {\rm ~km~sec}^{-1}$,
representing the escape speed from the parent cluster, are allowed to escape
through the reflective boundary.

\section{Results}\label{res}

\begin{table}
\centering
\caption{Summary of the computations for isolated clusters and those
with reflective boundary. The different columns represent as follows:
Col.~(1): identity of the particular model --- similar values with different names (ending with
A,B etc) imply runs repeated with different random seeds, Col.~(2): total number of stars $N(0)$ initially,
Col.~(3): number of computations $N_{sim}$ with the particular cluster,
Col.~(4): initial half-mass radius of the cluster $r_h(0)$ (isolated cluster) or radius of reflective sphere $R_s$,
Col.~(5): initial number of BHs $N_{BH}(0)$, Col.~(6): total number of BH-BH binary mergers within the
cluster $N_{mrg}$, Col.~(7): the times $t_{mrg}$ corresponding to the mergers, Col.~(8): number
of escaped BH-pairs $N_{esc}$ --- the three values of $N_{esc}$
are those with $T_{mrg}<3$ Gyr, 1 Gyr and 100 Myr respectively.}
{\small
\begin{tabular}{lllclllc}
\\
\hline
Model name &    $N(0)$   &  $N_{sim}$ & $r_h(0)$ or $R_s$ (pc)&  $N_{BH}(0)$ &    $N_{mrg}$   &   $t_{mrg}$ (Myr)  &  $N_{esc}$\\
\hline   
\multicolumn{8}{c}{Isolated clusters}\\
\hline
\vspace{-0.1 cm}
C10K20     & 10000   &   10       &   1.0              &   20         &       0        &  --- ---           &  --- ---  \\
\vspace{-0.1 cm}
C25K50     & 25000   &   10       &   1.0              &   50         &       0        &  --- ---           &  3 1 1    \\
\vspace{-0.1 cm}
C50K80     & 45000   &   1        &   1.0              &   80         &       1        &   698.3            &  3 1 0    \\
\vspace{-0.1 cm}
C50K80.1   & 45000   &   1        &   0.5              &   80         &       2        &   217.1, 236.6     &  3 2 1    \\ 
\vspace{-0.1 cm}
C50K40.1   & 45000   &   1        &   0.5              &   40         &       0        &   --- ---          &  1 1 1    \\  
\vspace{-0.1 cm}
C50K200    & 50000   &   1        &   1.0              &   200        &       2        &   100.8, 467.8     &  0 0 0    \\
\vspace{-0.1 cm}
C65K110    & 65000   &   1        &   1.0              &   110        &       1        &   314.6            &  4 2 1    \\
\vspace{-0.1 cm}
C65K110.1  & 65000   &   1        &   0.5              &   110        &       0        &  --- ---           &  4 3 1    \\
\vspace{-0.1 cm}
C65K55.1   & 65000   &   1        &   0.5              &   55         &       1        &   160.5            &  1 0 0    \\
\vspace{-0.1 cm}
C100K80    & 100000  &   1        &   1.0              &   80         &       2        &   219.4, 603.2     &  5 2 1    \\
\vspace{-0.1 cm}
C100K200   & 100000  &   1        &   1.0              &   200        &       0        &  --- ---           &  5 4 4    \\
\hline
\multicolumn{8}{c}{Reflective boundary}\\
\hline
\vspace{-0.1 cm}
R3K180     & 3000   &    1        &   0.4              &  180         &       1        &    1723.9          &  5 3 1    \\
\vspace{-0.1 cm}
R4K180A    & 4000   &    1        &   0.4              &  180         &       1        &    3008.8          &  2 2 1    \\
\vspace{-0.1 cm}
R4K180B    & 4000   &    1        &   0.4              &  180         &       2        &  100.2, 1966.5     &  2 1 0    \\
\vspace{-0.1 cm}
R3K100     & 3000   &    1        &   0.4              &  100         &       2        &  3052.8, 3645.9    &  1 1 0    \\
\vspace{-0.1 cm}
R4K100A    & 4000   &    1        &   0.4              &  100         &       2        &  104.4, 814.2      &  3 3 1    \\
\vspace{-0.1 cm}
R4K100B    & 4000   &    1        &   0.4              &  100         &       1        &    1135.3          &  3 3 3    \\
\hline
\end{tabular}
}
\label{tab1}
\end{table}

Table~\ref{tab1} summarizes the results of our computations.
For demonstration purposes, we choose one of the models, \viz, 
C50K80 (see Table~\ref{tab1}) --- other models generally
possess similar properties.
Fig.~\ref{fig:bhpos} (left panel) demonstrates the mass segregation of the BHs in the cluster which
takes about 50 Myr. As the BHs segregate within about $0.3 {\rm~pc}$ of the cluster core, the BH density
of becomes large enough to form BH-BH binaries through 3-body encounters.
Once BH-BH binaries begin forming, single BHs and BH binaries start escaping from the
BH-core due to the encounter-recoils (see Sec.~\ref{intro}). In Fig.~\ref{fig:bhpos} (left panel), one can
clearly distinguish the two phases of the BH subsystem --- the initial segregation
phase and the BH-core formation, the radial positions of the BHs being scattered outwards in the latter
phase due to the recoils. The corresponding decrease of $N_{BH}$ during this
phase is also shown in Fig.~\ref{fig:bhpos} (right panel).

\subsection{Mergers and escapers}\label{mrgesc}

\begin{figure}
\centering
\vspace{-2.2 cm}
\includegraphics[width=13.5 cm,angle=0]{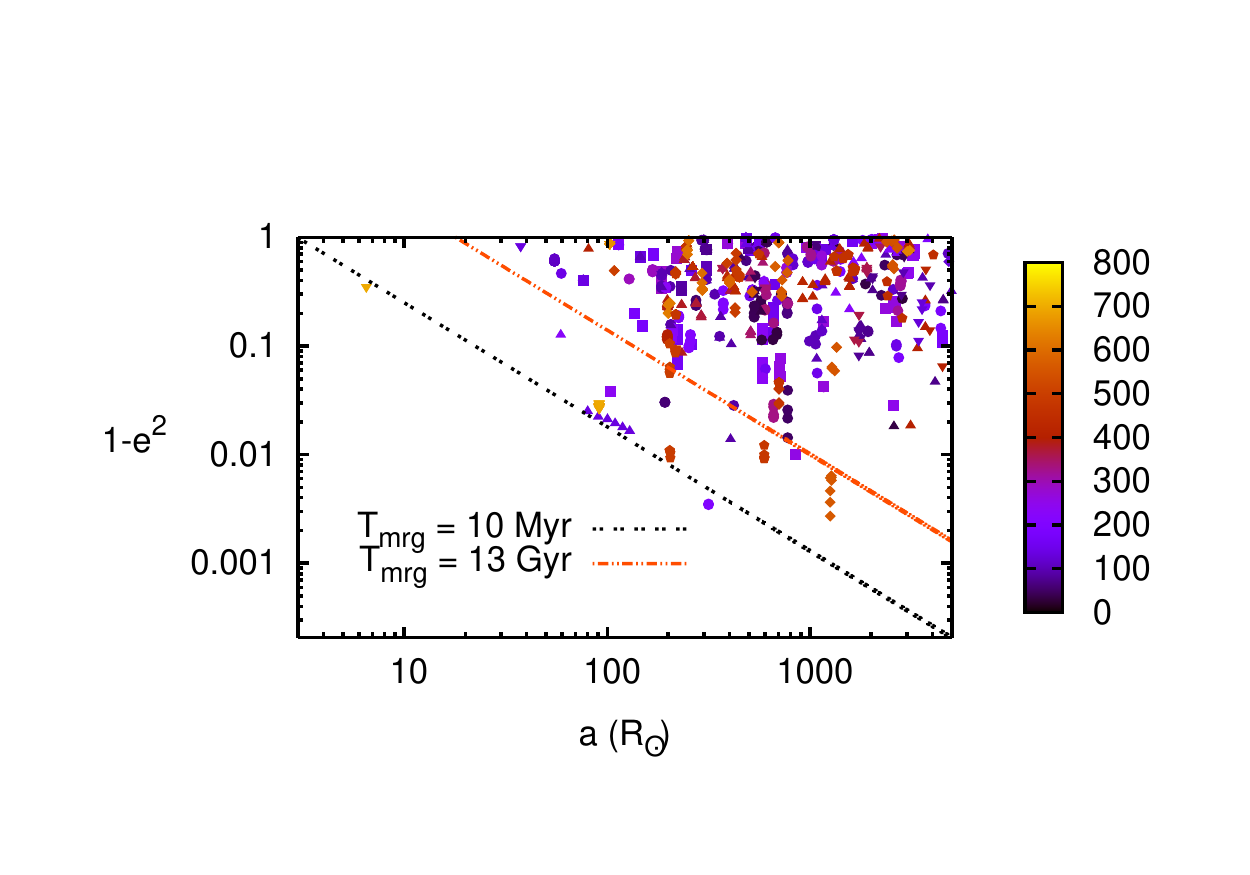}
\vspace{-1.3 cm}
\caption{The positions of all the BH binaries within the cluster on the $1-e^2$ vs. $a$
plane where the different symbols are used to distinguish between
different BH pairs (for model C50K80). The color-coding of the points, as indicated
by the color-scale, represents the time (in Myr)
at which they appear at a particular location in the
above plane. The lines of constant GW merger times, $T_{mrg}={\rm~} constant$,
are overplotted.}
\label{fig:aeplot}
\end{figure}

To study the possibility of BH-BH mergers, we consider the positions
of the BH-BH binaries within the cluster in an $a {\rm~ vs.~} (1-e^2)$ plane as shown in
Fig.~\ref{fig:aeplot}, where each pair is represented by a different symbol
and are color-coded with the cluster evolution time in Myr.
For each BH-pair, $a$ and $e$ fluctuate over the plane,
the changes occurring over a collision time-scale which is $\sim$ Myr. Since the orbital
periods of the corresponding binaries are much shorter
(from $\sim$ days to years), these points generally represent binaries which are stable
over many orbits. Overplotted in Fig.~\ref{fig:aeplot} are the lines of
constant GW merger time $T_{mrg}$ as given by Eqn.~(\ref{eq:tmrg}).
While most of the points correspond to very long merger times, a few of
them do lie close to the $T_{mrg}=10 {\rm ~Myr}$ line. This implies that
these binaries are indeed hardened up to small enough $a$ and/or acquire sufficient
eccentricity that had they been left unperturbed, they would have merged via GW emission
within several Myr. However, these merging BH-pairs can still be perturbed by further
encounters on timescales $\sim$ Myr which can
often prevent them from merging. In the particular example shown in
Fig.~\ref{fig:aeplot}, only one of them could merge within the cluster.
This feature is found to be generally true for all the models reported here (see Table~\ref{tab1}),
\ie, each of them does produce BH-pairs that are capable of merging within
several Myr. On the other hand, among the escaped BH binaries, all of them with GW merger times smaller
than a Hubble time are of interest since they remain unperturbed afterwards. For each of our computed
models, a few of the escaped BH-BH binaries can merge in $T_{mrg} < 3$ Gyr as can be read from
Table \ref{tab1} (also, see reference \cite{bbk2010}).
We also find that for all of the computed models, most mergers occur
within a few Gyr (see reference \cite{bbk2010}).

The above results imply that an important class of star clusters for dynamically
forming BH binaries that merge at the present epoch are those
with initial mass $M_{cl}(0) \ga 3\times 10^4 \Ms$ and are few Gyr old.
Such clusters represent intermediate-age massive clusters (hereafter IMC)
with initial masses close to the upper-limit of
the initial cluster mass function (ICMF) in spiral and starburst
galaxies \cite{lrs2009}.
Globular clusters, which are at least an order of magnitude more massive,
are typically much older ($\sim 10 {\rm ~Gyr}$),
so that they cannot contribute significantly to the present-day merger rate, as most
of the mergers from them would have occurred earlier. On the other hand, young
massive clusters, with ages
$ < 50$ Myr, are generally too young to produce BH-BH mergers, as the stratification
of the BHs and the formation of the BH core itself takes longer.
Hence, IMCs seem to be the most likely star clusters for dynamically generating
present-day BH-BH merger events.

\subsection{Detection rate of BH-BH mergers}\label{detrate}

The results from Table~\ref{tab1} give an average merger rate of $\approx 0.4$/cluster/Gyr
for those occurring within the clusters and
$\approx 0.9$/cluster/Gyr for the escapers, totaling ${\mathcal R}_{mrg}\approx 1.3$/cluster/Gyr
(see reference \cite{bbk2010}). To estimate the detection rate of BH-BH
mergers from IMCs, we adopt the space density, as derived by reference \cite{pzm2000},
for young populous clusters which is $\rho_{cl} \approx 3.5 {\rm ~} h^3 {\rm ~Mpc}^{-3}$, where
$h=0.73$ is the Hubble parameter. This merger rate implies that
the IMCs would yield a BH-BH merger detection
rate of ${\mathcal R}_{\rm AdLIGO} \approx 30 {\rm ~yr}^{-1}$ for the upcoming
``Advanced LIGO'' (AdLIGO) GW detector (see reference \cite{bbk2010} and citations therein).
Interestingly, the above dynamical BH-BH merger detection rate can be more than
10 times higher than that from individual primordial binaries \cite{bky2007}.
Our results then imply that the dynamically formed BH-BH binaries constitute
the dominant contribution to the BH-BH merger detection from the Universe \cite{bbk2010}.

\section{Introduction of stellar evolution}\label{evolintro}

Although the study described above has provided significant insights into the dynamics
of stellar mass BHs in star clusters and its role in dynamically generating BH-BH merger
events and has been performed self-consistently using direct N-body computations, the
cluster models would be more realistic if the BHs are formed with a mass function
from stellar evolution. To materialize such improvements, we model the 
initial Plummer clusters with all stars at their Zero-Age-Main-Sequence (ZAMS) having
a Kroupa IMF \cite{krp2001} (upto $100\Ms$). These models are also evolved using the NBODY6 integrator
where the direct N-body integration of the cluster stars is coupled with their stellar
evolution using analytic but well tested stellar evolution recipes \cite{hur2000}.

\begin{figure}[!ht]
\centering
\includegraphics[width=7.4cm,angle=0]{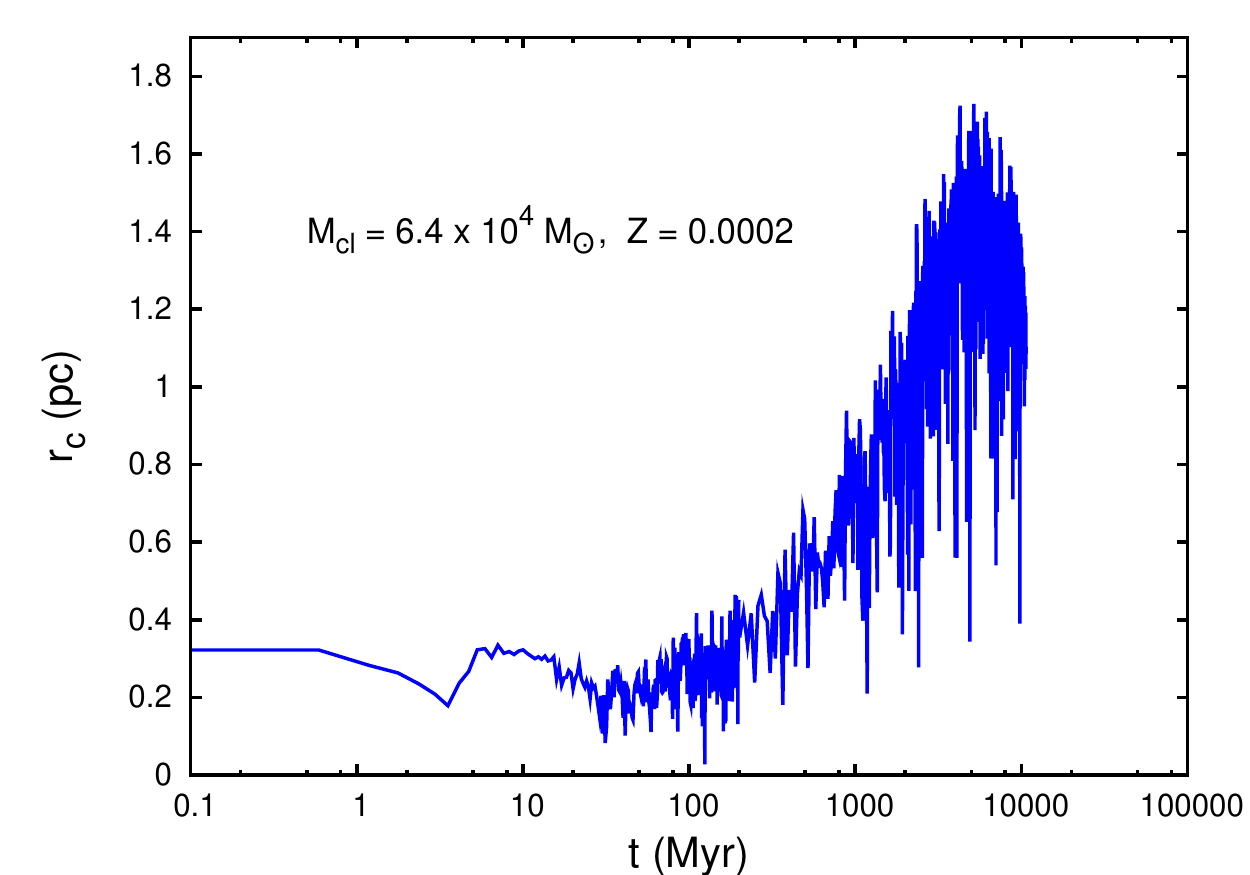}
\includegraphics[width=7.4cm,angle=0]{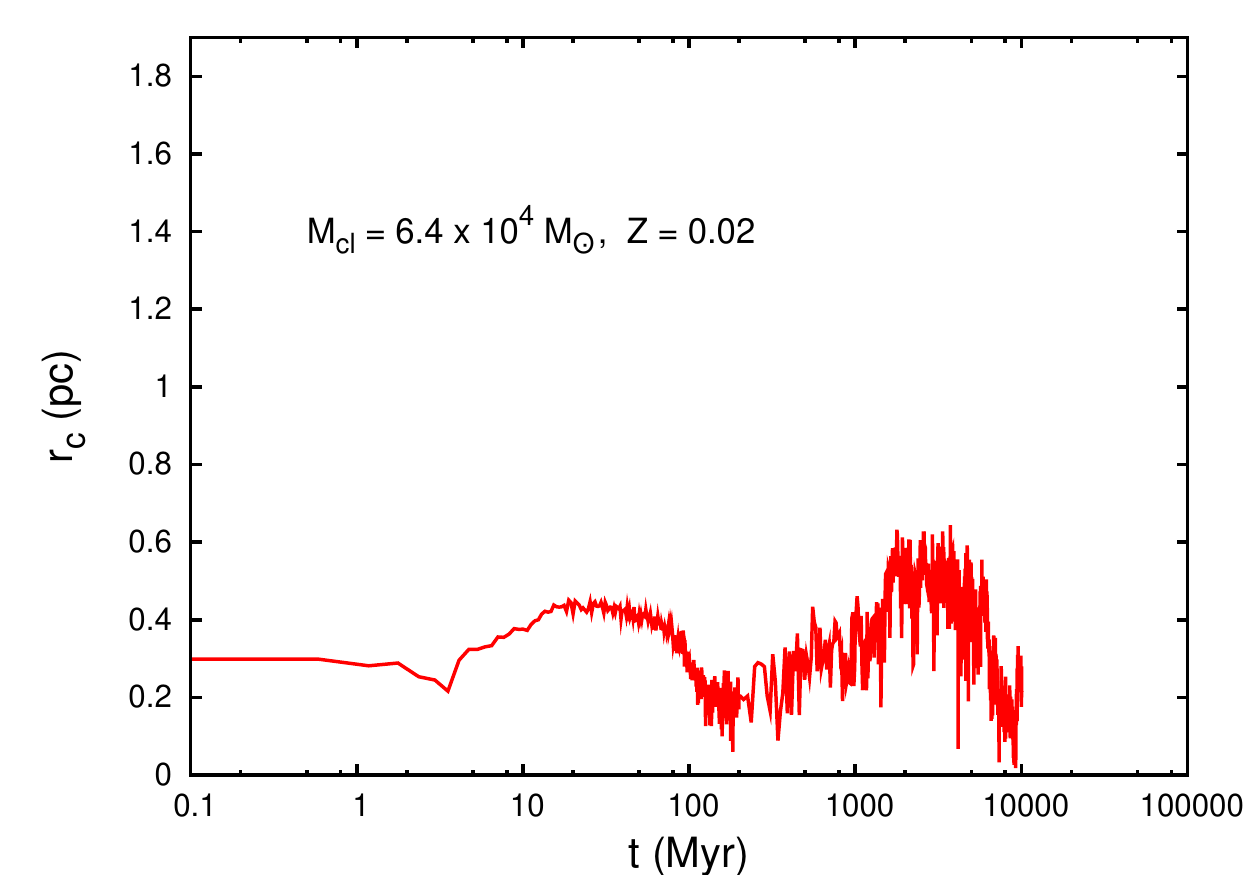}
\caption{\small Evolution of core radii $r_c$ for two computed clusters with metallicities
$Z=Z_\odot$ and $Z=0.01Z_\odot$ but otherwise identical initial conditions (see text).
A full BH retention is assumed.
Such comparison demonstrates that the heating effect of the BH core and the resulting expansion
of $r_c$ is significantly larger for the lower metallicity cluster (left panel)
as the BHs formed are considerably more massive in that case.}
\label{fig:corecomp}
\end{figure}

Fig.~\ref{fig:corecomp} shows the evolution of the core radii $r_c$ for two cluster models
with different metallicities, \viz, $Z=Z_\odot$ and $Z=0.01Z_\odot$,
having otherwise exactly the same initial conditions ($N(0) = 10^5$, $r_h(0) = 1$ pc Plummer
spheres and all stars having identical positions, velocities and masses following a Kroupa IMF).
All the BHs are retained in the clusters following their formation via supernovae.
For the cluster with lower $Z$, the BH masses are considerably larger than
that for the higher $Z$ case (average BH mass $\approx 14\Ms$ for $Z=0.01Z_\odot$
as opposed to $\approx 7\Ms$ for $Z=Z_\odot$) so that the expansion of the core radius
due to the heating effect of the BH sub-cluster (see Sec.~\ref{intro}) is significantly
more prominent for the lower $Z$ model, as can be seen in Fig.~\ref{fig:corecomp}. 
Also, the $Z=0.01Z_\odot$ model have produced 3 BH-BH binaries among its escapers
that can merge via GW radiation within a Hubble time while its high $Z$ counterpart
have produced only one. This can be expected as the heavier BHs in the lower $Z$ cluster
create a denser BH core where the binary BH formation rate through 3-body encounters
and their subsequent dynamical hardening rates are higher. This indicates that
the modification of the dynamical evolution of a cluster by its BHs and the dynamical
BH-BH merger rate depend crucially on its metallicity. Studying the effect of the
stellar content of a cluster and of its properties on the dynamics of its BHs
is under progress.    

\section*{Acknowledgment}

I thank Holger Baumgardt (Univ. Queensland, Australia) and Pavel Kroupa (Univ. Bonn, Germany)
for close collaboration regarding this work. I take this opportunity to
thank the organizers of the 25th Texas Symposium, 2010 event and the session chairs for selecting
my contribution. I thank the Alexander von Humboldt foundation for support during this work.

\end{document}